\documentclass[longauth]{aa}  
\usepackage{graphicx}
\usepackage{txfonts}
\begin{document} 

\title{The LOFAR long baseline snapshot calibrator survey}


\author{
J.~Mold\'{o}n\inst{1}\and 
A.~T.~Deller\inst{1}\and 
O.~Wucknitz\inst{2}\and 
N.~Jackson\inst{3}\and 
A.~Drabent\inst{4}\and 
T.~Carozzi\inst{5}\and 
J.~Conway\inst{5}\and 
A.~D.~Kapi\'{n}ska\inst{6,7,8}\and 
J.~P.~McKean\inst{1}\and 
L.~Morabito\inst{9}\and 
E.~Varenius\inst{5}\and 
P.~Zarka\inst{10}\and 
J.~Anderson\inst{11}\and 
A.~Asgekar\inst{1,12}\and 
I.~M.~Avruch\inst{13,14}\and 
M.~E.~Bell\inst{15}\and 
M.~J.~Bentum\inst{1,16}\and 
G.~Bernardi\inst{17}\and 
P.~Best\inst{18}\and 
L.~B\^{i}rzan\inst{9}\and 
J.~Bregman\inst{1}\and 
F.~Breitling\inst{19}\and 
J.~W.~Broderick\inst{20,21}\and 
M.~Br\"uggen\inst{22}\and 
H.~R.~Butcher\inst{23}\and 
D.~Carbone\inst{24}\and 
B.~Ciardi\inst{25}\and 
F.~de Gasperin\inst{22}\and 
E.~de Geus\inst{1,26}\and 
S.~Duscha\inst{1}\and 
J.~Eisl\"offel\inst{4}\and 
D.~Engels\inst{27}\and 
H.~Falcke\inst{28,1}\and 
R.~A.~Fallows\inst{1}\and 
R.~Fender\inst{20}\and 
C.~Ferrari\inst{29}\and 
W.~Frieswijk\inst{1}\and 
M.~A.~Garrett\inst{1,9}\and 
J.~Grie\ss{}meier\inst{30,31}\and 
A.~W.~Gunst\inst{1}\and 
J.~P.~Hamaker\inst{1}\and 
T.~E.~Hassall\inst{21}\and 
G.~Heald\inst{1}\and 
M.~Hoeft\inst{4}\and 
E.~Juette\inst{32}\and 
A.~Karastergiou\inst{20}\and 
V.~I.~Kondratiev\inst{1,33}\and 
M.~Kramer\inst{2,3}\and 
M.~Kuniyoshi\inst{2}\and 
G.~Kuper\inst{1}\and 
P.~Maat\inst{1}\and 
G.~Mann\inst{19}\and 
S.~Markoff\inst{24}\and 
R. McFadden\inst{1}\and 
D.~McKay-Bukowski\inst{34,35}\and 
R.~Morganti\inst{1,14}\and 
H.~Munk\inst{1}\and 
M.~J.~Norden\inst{1}\and 
A.~R.~Offringa\inst{23}\and 
E.~Orru\inst{1}\and 
H.~Paas\inst{36}\and 
M.~Pandey-Pommier\inst{37}\and 
R.~Pizzo\inst{1}\and 
A.~G.~Polatidis\inst{1}\and 
W.~Reich\inst{2}\and 
H.~R\"ottgering\inst{9}\and 
A.~Rowlinson\inst{15}\and 
A.~M.~M.~Scaife\inst{21}\and 
D.~Schwarz\inst{38}\and 
J.~Sluman\inst{1}\and 
O.~Smirnov\inst{39,40}\and 
B.~W.~Stappers\inst{3}\and 
M.~Steinmetz\inst{19}\and 
M.~Tagger\inst{30}\and 
Y.~Tang\inst{1}\and 
C.~Tasse\inst{41}\and 
S.~Thoudam\inst{28}\and 
M.~C.~Toribio\inst{1}\and 
R.~Vermeulen\inst{1}\and 
C.~Vocks\inst{19}\and 
R.~J.~van Weeren\inst{17}\and 
S.~White\inst{25}\and 
M.~W.~Wise\inst{1,24}\and 
S.~Yatawatta\inst{1}\and 
A.~Zensus\inst{2}
}
\institute{
ASTRON, the Netherlands Institute for Radio Astronomy, Postbus 2, 7990 AA, Dwingeloo, The Netherlands \and
Max-Planck-Institut f\"{u}r Radioastronomie, Auf dem H\"ugel 69, 53121 Bonn, Germany \and
Jodrell Bank Center for Astrophysics, School of Physics and Astronomy, The University of Manchester, Manchester M13 9PL,UK \and
Th\"{u}ringer Landessternwarte, Sternwarte 5, D-07778 Tautenburg, Germany \and
Onsala Space Observatory, Dept. of Earth and Space Sciences, Chalmers University of Technology, SE-43992 Onsala, Sweden \and
Institute of Cosmology \& Gravitation, University of Portsmouth, Burnaby Road, PO1 3FX Portsmouth, UK \and
ARC Centre of Excellence for All-sky astrophysics (CAASTRO), Sydney Institute of Astronomy, University of Sydney Australia \and
International Centre for Radio Astronomy Research - Curtin University, GPO Box U1987, Perth, WA 6845, Australia \and
Leiden Observatory, Leiden University, PO Box 9513, 2300 RA Leiden, The Netherlands \and
LESIA-Observatoire de Paris, CNRS, UPMC Univ Paris 6, Univ. Paris-Diderot, France \and
Helmholtz-Zentrum Potsdam, DeutschesGeoForschungsZentrum GFZ, Department 1: Geodesy and Remote Sensing, Telegrafenberg, A17, 14473 Potsdam, Germany \and
Shell Technology Center, Bangalore, India \and
SRON Netherlands Insitute for Space Research, PO Box 800, 9700 AV Groningen, The Netherlands \and
Kapteyn Astronomical Institute, PO Box 800, 9700 AV Groningen, The Netherlands \and
CSIRO Australia Telescope National Facility, PO Box 76, Epping NSW 1710, Australia \and
University of Twente, The Netherlands \and
Harvard-Smithsonian Center for Astrophysics, 60 Garden Street, Cambridge, MA 02138, USA \and
Institute for Astronomy, University of Edinburgh, Royal Observatory of Edinburgh, Blackford Hill, Edinburgh EH9 3HJ, UK \and
Leibniz-Institut f\"{u}r Astrophysik Potsdam (AIP), An der Sternwarte 16, 14482 Potsdam, Germany \and
Astrophysics, University of Oxford, Denys Wilkinson Building, Keble Road, Oxford OX1 3RH \and
School of Physics and Astronomy, University of Southampton, Southampton, SO17 1BJ, UK \and
University of Hamburg, Gojenbergsweg 112, 21029 Hamburg, Germany \and
Research School of Astronomy and Astrophysics, Australian National University, Mt Stromlo Obs., via Cotter Road, Weston, A.C.T. 2611, Australia \and
Anton Pannekoek Institute, University of Amsterdam, Postbus 94249, 1090 GE Amsterdam, The Netherlands \and
Max Planck Institute for Astrophysics, Karl Schwarzschild Str. 1, 85741 Garching, Germany \and
SmarterVision BV, Oostersingel 5, 9401 JX Assen \and
Hamburger Sternwarte, Gojenbergsweg 112, D-21029 Hamburg \and
Department of Astrophysics/IMAPP, Radboud University Nijmegen, P.O. Box 9010, 6500 GL Nijmegen, The Netherlands \and
Laboratoire Lagrange, UMR7293, Universit\`{e} de Nice Sophia-Antipolis, CNRS, Observatoire de la C\'{o}te d'Azur, 06300 Nice, France \and
LPC2E - Universite d'Orleans/CNRS \and
Station de Radioastronomie de Nancay, Observatoire de Paris - CNRS/INSU, USR 704 - Univ. Orleans, OSUC , route de Souesmes, 18330 Nancay, France \and
Astronomisches Institut der Ruhr-Universit\"{a}t Bochum, Universitaetsstrasse 150, 44780 Bochum, Germany \and
Astro Space Center of the Lebedev Physical Institute, Profsoyuznaya str. 84/32, Moscow 117997, Russia \and
Sodankyl\"{a} Geophysical Observatory, University of Oulu, T\"{a}htel\"{a}ntie 62, 99600 Sodankyl\"{a}, Finland \and
STFC Rutherford Appleton Laboratory,  Harwell Science and Innovation Campus,  Didcot  OX11 0QX, UK \and
Center for Information Technology (CIT), University of Groningen, The Netherlands \and
Centre de Recherche Astrophysique de Lyon, Observatoire de Lyon, 9 av Charles Andr\'{e}, 69561 Saint Genis Laval Cedex, France \and
Fakult\"{a}t f\"{u}r Physik, Universit\"{a}t Bielefeld, Postfach 100131, D-33501, Bielefeld, Germany \and
Department of Physics and Elelctronics, Rhodes University, PO Box 94, Grahamstown 6140, South Africa \and
SKA South Africa, 3rd Floor, The Park, Park Road, Pinelands, 7405, South Africa \and
LESIA, UMR CNRS 8109, Observatoire de Paris, 92195   Meudon, France 
}

   \date{Received XXXX}

 
  \abstract
{}
{An efficient means of locating calibrator sources for International LOFAR is
developed and used to determine the average density of usable calibrator
sources on the sky for subarcsecond observations at 140 MHz.}
{We used the multi-beaming capability of LOFAR to conduct a fast and
computationally inexpensive survey with the full International LOFAR array.
Sources were pre-selected on the basis of 325~MHz arcminute-scale flux density
using existing catalogues. By observing 30 different sources in each of the 12
sets of pointings per hour, we were able to inspect 630 sources in two hours to
determine if they possess a sufficiently bright compact component to be usable
as LOFAR delay calibrators.}
{Over 40\% of the observed sources are detected on multiple baselines between
international stations and 86 are classified as satisfactory calibrators.  We
show that a flat low-frequency spectrum (from 74 to 325~MHz) is the best
predictor of compactness at 140~MHz. We extrapolate from our sample to show
that the density of calibrators on the sky that are sufficiently bright to
calibrate dispersive and non-dispersive delays for the International LOFAR
using existing methods is 1.0 per square degree.}
{The observed density of satisfactory delay calibrator sources means that
observations with International LOFAR should be possible at virtually any point
in the sky, provided that a fast and efficient search using the methodology
described here is conducted prior to the observation to identify the best
calibrator.}


   \maketitle
%

\section{Introduction}

High angular resolution (subarcsecond) observations at long wavelengths
($\lambda >$1~m) can be used for a wide variety of astronomical applications.
Examples include measuring the angular broadening of galactic objects due to
interstellar scattering, spatially localising low-frequency emission identified
from low-resolution observations, extending the wavelength coverage of studies
of (for instance) Active Galactic Nuclei (AGN) at a matched spatial resolution,
or studying the evolution of black holes throughout the universe by means of
high-resolution low-frequency surveys \citep{falke04}.  However, such
observations have only rarely been employed in the past, due to the difficulty
of calibrating the large, rapid delay and phase fluctuations induced by the
differential ionosphere seen by widely separated stations.  Shortly after the
first Very Long Baseline Interferometry (VLBI) observations at cm wavelengths,
similar observations were performed on a number of bright radio Active Galactic
Nuclei (AGN) and several strong, nearby radio pulsars
\citep{clark75a,vandenberg76a} at frequencies of 74--196~MHz with baselines of
up to 2500 km (providing angular resolution as high as 0.12\arcsec). More
recently, \cite{nigl07} conducted 20~MHz VLBI observations on a single baseline
between Nan\c{c}cay and the LOFAR's Initial Test Station on Jupiter bursts.
These early efforts were limited to producing size estimates (or upper limits)
using the visibility amplitude information; imaging was not performed.  Before
the construction of the Low Frequency Array (LOFAR; \citealt{van-haarlem13a}),
the lowest frequency at which subarcsecond imaging has been performed is
325~MHz \citep[e.g.,][]{wrobel86a,ananthakrishnan89a,lenc08a}.

With the commissioning of LOFAR, true subarcsecond imaging at frequencies below
300~MHz is now possible for the first time.  With a current maximum baseline of
1300~km, the International LOFAR array is capable of attaining an angular
resolution of $\sim$0.4\arcsec\ at a frequency of 140 MHz.  The high
sensitivity and wide bandwidth coverage of LOFAR, coupled with advances in
electronic stability, greatly mitigate the issues faced by the early
low-frequency efforts.

The early attempts of using international baselines of LOFAR are described by
\citet{wucknitz10a}.  This includes the first-ever long-baseline LOFAR images
that were produced of 3C196 in the low band (30-80 MHz) with a resolution of
about one arcsec (more than an order of magnitude better than previously
possible) using only a fraction of the final array. Following these first
experiments, a number of calibration strategies were tested, finding that after
conversion to a circular polarisation basis (as described in
Section~\ref{sec:datareduction}) standard VLBI calibration approaches are
sufficient to correct for the large and rapidly varying dispersive delay
introduced by the differential ionosphere above each station within narrow
frequency bands. This implies that imaging of small fields around bright
compact sources is relatively straightforward.  The calibration of visibility
amplitudes still requires significant effort; this is discussed further in
Section~\ref{sec:analysis}. More recently, the first high band (110--160~MHz)
observation with the LOFAR long baselines was presented in Varenius et al.
(A\&A, submitted), where subarcsecond images of M82 were presented.

Imaging of faint sources -- where calibration solutions cannot be directly
derived -- remains challenging, as large spatial gradients in the dispersive
ionospheric delay severely limit the area over which a calibration solution can
be extrapolated.  At cm-wavelengths, it is common VLBI practice to make use of
a calibrator at a separation up to $\sim$5 degrees \citep[e.g.,][]{walker99a}
to solve the gradient in phase across the observing band ({\em delay}), the
phase at the band centre ({\em phase}) and the rate of change of phase at the
band centre with time ({\em rate}) with a solution interval of minutes.  With
over $\sim$7600 VLBI calibrators now
known\footnote{http://astrogeo.org/vlbi/solutions/rfc\_2014c/}, with a density
of ~$\sim$0.2 per square degree, almost any target direction can find a
suitable calibrator at cm wavelengths.  At metre wavelengths, however, a given
change in Total Electron Content (TEC) has a much larger impact on the delay,
phase and rate, as discussed in Section~\ref{sec:analysis}.  This means that
much smaller spatial extrapolations can be tolerated before unacceptably large
residual errors are seen.  Moreover, many of the known cm-VLBI calibrators have
inverted spectra or a low-frequency turnover, making them insufficiently bright
at LOFAR wavelengths. 

Accordingly, identifying sufficiently bright and compact sources to use as
calibrators is of the utmost importance for the general case of observing with
International LOFAR at the highest resolutions.  Unsurprisingly, however, very
little is known about the compact source population at this frequency range.
\citet{lenc08a} used global VLBI observations to study the compact source
population in large fields around the gravitational lens B0218+357 and a nearby
calibrator at 325~MHz, by imaging sources selected from lower resolution
catalogues. They found that about 10\% of candidate sources brighter than
$\sim$100 mJy at 325~MHz could be detected at $\sim0.1$~arcsecond resolution.
Based on this, \citet{lenc08a} estimated a density of compact sources above
10~mJy at 240~MHz of 3 deg$^{-2}$.  Later, \citet{wucknitz10b} applied an
efficient wide-field mapping method to image the entire primary beam for one of
the fields, finding exactly the same sources.  \cite{rampadarath09a} analysed
archival Very Long Baseline Array (VLBA) observations of 43 sources at 325~MHz,
finding 30 which would be satisfactory calibrators for International LOFAR
observations, but were not able to draw any conclusions about the density of
satisfactory calibrators in general.

In this paper, we present results of LOFAR commissioning observations which
targeted 720 radio sources at high angular resolution in two hours of observing
time (the ``LOFAR snapshot calibrator survey'').  We show that the observing
and data reduction strategy employed is a robust and efficient means to
identify suitable bright (``primary'') calibrators prior to a normal LOFAR
science observation.  By analysing the results in hand, we estimate the density
of suitable primary calibrators for International LOFAR on the sky. Finally, we
propose an efficient procedure to search and identify all the necessary
calibrators for any given International LOFAR observation, which can be
undertaken shortly before a science observation. In Appendix~\ref{apx:plan}, we
give a procedure which can be followed to set up an observation with
International LOFAR, using the tools developed in this work.

\section{Calibration of International LOFAR observations} 
\label{sec:calibration}

The majority of the LOFAR stations, namely the core and remote stations, are
distributed over an area roughly 180 km in diameter predominantly in the
northeastern Dutch province of Drenthe. Currently, the array also includes 8
international LOFAR stations across Europe that provide maximum baselines up to
1292~km. One additional station is planned to be completed in Hamburg (Germany)
in 2014, and three stations in Poland will commence construction in 2014,
extending the maximum baseline to $\sim$2000~km.
Table~\ref{tab:ilofarstations} shows the distance from each current
international LOFAR station to the LOFAR core, and the corresponding resolution
provided by the international station to core baseline at 140 MHz\footnote{An
up-to-date map of all LOFAR stations can be found at
\url{http://www.astron.nl/~heald/lofarStatusMap.html}.}.  

\begin{table}   
\caption{Current international LOFAR stations}
\label{tab:ilofarstations}
\centering
\begin{tabular}{l c c} 
\hline\hline
Station & Distance to  &  Corresponding resolution  \\
            & LOFAR core (km) &  at 140 MHz (\arcsec) \\
\hline
DE605  & 226  &  2.4  \\
DE601  & 266  &  2.0  \\
DE603  & 396  &  1.4  \\
DE604  & 419  &  1.3  \\
DE602  & 581  &  0.9  \\
SE607  & 594  &  0.9  \\
UK608  & 602  &  0.9  \\
FR606  & 700  &  0.8  \\
\hline
\end{tabular}
\end{table}

Calibration of these long baselines poses a special challenge compared to LOFAR
observations with the Dutch array, and these can be addressed using tools
developed for cm~wavelength VLBI.  The calibration process must derive the
station-based amplitude and phase corrections in the direction of the target
source with adequate accuracy as a function of time.  Amplitude corrections are
generally more stable with time and sky offset, and the process differs little
from shorter baseline LOFAR observations (aside from the problems of first
deriving a reasonable model of the calibrator source, which is discussed in
Section~\ref{sec:analysis}), so we do not discuss amplitude calibration here.
Below, we first define some VLBI terminology and briefly describe phase
calibration in cm~VLBI, before describing the adaptations necessary for LOFAR.

\subsection{VLBI calibration at cm wavelengths}

Due to the large and time-variable delay offsets at each station, solving for
phase corrections directly (approximating the correction as constant over a
given solution time and bandwidth) would require very narrow solution intervals
for VLBI, and hence an extremely bright calibrator source.  However, such a
source would be unlikely to be close on the sky to the target, with a
separation of perhaps tens of degrees, and the differential
atmosphere/ionosphere between the calibrator and the target direction would
render the derived calibration useless in the target direction.  In order to
make use of calibrators closer to the target, VLBI calibration therefore solves
for 3 parameters (phase, non-dispersive delay, and phase rate) in each solution
interval, allowing the solution duration and bandwidth to be greatly extended.
This approach makes a number of assumptions:

\begin{enumerate}
\item The change of the delay resulting from the dispersion is negligible,
so the total delay can be approximated as a constant across the solution
bandwidth;
\item The change in delay over the solution time can be approximated in a linear
fashion;
\item The change in phase over the solution bandwidth due to the change in delay
over the solution time is small (since the time variation is approximated with a
phase rate, rather than a delay rate).
\end{enumerate}

Meeting these assumptions requires that the solution interval and bandwidth be
kept relatively small, which is at odds with the desire to maximise
sensitivity, which demands that the solution interval and bandwidth be as large
as possible.

\begin{figure*}[th!!!] 
\center
\includegraphics[angle=0]{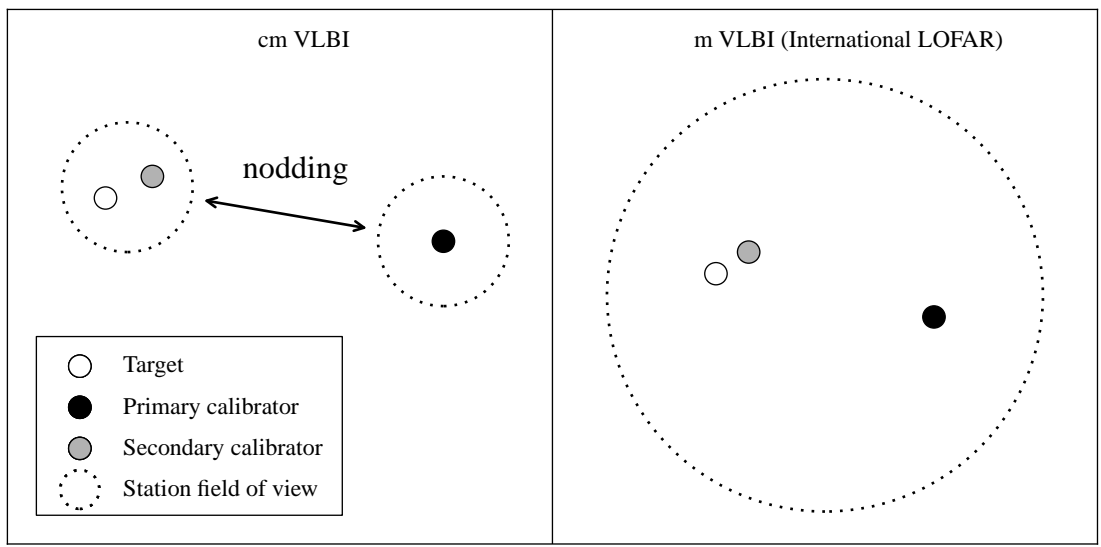}
\caption{Typical calibration setup for cm VLBI (left) and International LOFAR
(right).  Note that in some cases the target may itself function as the
secondary calibrator.  A secondary calibrator is not always required for cm
VLBI, but will almost always be needed for International LOFAR, unless the
primary calibrator is fortuitously close.  The larger field of view of LOFAR
means that both the primary and secondary calibrators will always be observed
contemporaneously, unlike in cm VLBI, where nodding between the primary
calibrator and target is typically required (shown by the double arrow in the
left panel).}
\label{fig:calstrategy}
\end{figure*}

For cm VLBI, the dispersive delay due to the ionosphere is small (and so are
the changes with time), meaning that solution intervals of duration minutes and
width tens to hundreds of MHz are generally permissible.  After application of
the solutions from the primary calibrator, it is common to use a secondary
calibrator\footnote{A secondary calibrator is often referred to as an
``in-beam'' calibrator if it is close enough to the target source to be
observed contemporaneously} closer to the target source (separation
$\sim$arcmin), or to use the target source if it is bright enough for
``self-calibration'', solving only for the phase (no delay or rate). This
second phase-only calibration is used to refine the calibration errors that
result from the spatial or temporal interpolation of the primary solutions.
Because this is a problem with fewer degrees of freedom, lower signal-to-noise
ratio (S/N) data can be used.  Additionally, because the bulk delay has already
been removed, even more bandwidth can be combined in a single solution for a
further improvement in S/N.  A secondary calibrator can therefore be
considerably fainter (usually $\sim$1-10 mJy versus $>$100 mJy for a primary
calibrator).  This typical VLBI calibration strategy is illustrated in
Figure~\ref{fig:calstrategy}.

To reiterate: in standard VLBI, a {\em primary} calibrator is used to solve for
the bulk delay and rate offsets in the approximate direction of the target
source.  Usually this primary calibrator cannot be observed contemporaneously
with the target source, and so a nodding calibration is used, where scans on
the target are interleaved between scans on the calibrator.  Depending on the
observing frequency and conditions, and the separation to the primary
calibrator, no further refinement may be needed.  However, it is common to use
a {\em secondary} calibrator, or the target itself if bright enough, to derive
further phase-only corrections.  Naturally, the use of phase-only corrections
imposes the requirement that the differential delay between the primary
calibrator direction and the target direction be small enough that it can be
approximated as a constant phase offset across the width of the primary
solution bandwidth.  For bandwidths of tens of MHz, this means the differential
delay error must be $\lesssim$1 ns. Finally, all of the observing bandwidth can
be combined to derive these secondary corrections, improving sensitivity.  

\subsection{Application to LOFAR}

For 110--240~MHz (LOFAR High Band) observations on long baselines, the
approximations made when solving for phase, phase rate, and non-dispersive
delay fail badly when applied to bandwidths of tens of MHz or more.  Two
options present themselves: to add additional parameters (covering dispersive
delay and dispersive delay rate) to the global fit, or to reduce the solution
bandwidth such that the constant dispersive delay approximation becomes valid
again.  The former option is obviously preferable from a sensitivity
perspective, but greatly expands and complicates the solution search space.
Efforts are underway to implement such an expanded fit, including in addition
differential Faraday rotation, which becomes increasingly important at
frequencies below 100 MHz.  First tests on individual long baselines of LOFAR
as well as baselines to other telescopes are promising, but the algorithms are
not yet sufficiently mature for automatic calibration.  Accordingly, we focus
here on sources which can serve as primary calibrators under the latter set of
conditions, where solution bandwidths are limited to no more than a few MHz.

The system equivalent flux density (SEFD) of a single LOFAR core station is
approximately 1500~Jy\footnote{A LOFAR core station consists of two
sub-stations ($2\times24$ tiles) in the HBA.} at a frequency of $\sim$140~MHz
\citep{van-haarlem13a}.  An international station has twice the collecting area
of a core station at $\sim$140~MHz, so the expected SEFD is around 750 Jy.  The
24 core stations can be coherently combined into a single phased array with an
SEFD of $\sim$65~Jy, in the absence of correlated noise (i.e., when the
observed field only contains sources significantly fainter than the station
SEFD).  The theoretical 1$\sigma$ baseline sensitivity of an international
station to the phased-up core station, given 3~MHz of bandwidth and 4 minutes
of observing time, is hence 8~mJy in a single polarisation.  A source with a
compact flux density of 50~mJy yields a theoretical baseline signal-to-noise
ratio of 6, and is therefore a potential primary calibrator. In the real world,
the sensitivity of the phased-up core station will be reduced by failing tiles,
imperfect calibration and correlated (astronomical) noise, and so 50~mJy should
be considered a lower limit on the useful primary calibrator flux density.

In addition to being sufficiently bright, the primary calibrator must be close
enough to the secondary calibrator/target field that the differential delay
between the two fields does not lead to decorrelation when phase-only secondary
calibration is performed (just as for cm~VLBI).  The solution bandwidths are
narrower by a factor of $\gtrsim$10 than for cm~VLBI, which is helpful, but the
ionospheric delay (inversely proportional to observing frequency squared) is
much greater, meaning that on balance a closer calibrator will be needed than
the $\lesssim$5 degrees typical for cm~VLBI.  The maximum acceptable separation
will be a strong function of ionospheric conditions and elevation, but at face
value, given a bandwidth 20 times narrower (e.g., 3 MHz vs 64 MHz) and
frequency 10 times lower (140 MHz vs 1400 MHz), one would expect that the
calibrator would need to be separated by $\lesssim$1 degree.  This is borne out
by commissioning observations with LOFAR, which have shown acceptable results
with separations up to several degrees in favourable ionospheric conditions,
and unacceptable results with separations as small as $\sim$0.8 degrees in poor
conditions.  Ideally, then, a primary calibrator for International LOFAR
observations would be located $\lesssim$1 degree from the secondary
calibrator/target field to give acceptable calibration under most
circumstances.  As illustrated in Figure~\ref{fig:calstrategy}, this leads to
the one calibration advantage of International LOFAR compared to cm~VLBI; since
the beam of an International LOFAR station is $\gtrsim$2 degrees across, the
primary calibrator will by necessity be observed contemporaneously with the
target source.

This paper focuses on the identification of {\em primary} calibrators for High
Band (110--240 MHz) International LOFAR observations.  The reason is that after
primary calibration, the bandwidth can be increased by a factor of $\sim$30 for
secondary calibration. That means that for target sources brighter than
$\sim$10~mJy the target itself can serve as secondary calibrator. Even when the
target is not sufficiently bright, the density of these faint sources on the
sky is high enough that a suitable calibrator should be found very close to the
target.  We discuss the identification of secondary calibrators further in
Section~\ref{sec:skydensity}.

\section{Observations and data reduction} \label{sec:obs}

When operating in 8-bit mode \citep{van-haarlem13a}, LOFAR has 488 sub-bands of
width 0.195 kHz which can be flexibly distributed over a number of beams. For
our purposes a potential calibrator source must be detected within a single
3~MHz band to be useful, so we could divide the available bandwidth over a large
number of target sources, enabling rapid surveying.

Two hour-long commissioning observations were conducted in May and November
2013, as summarised in Table~\ref{tab:obs}.  All available LOFAR stations were
utilised: 24 core stations and 13 remote stations in the Netherlands, and 8
international stations (see \cite{van-haarlem13a} for the list and locations of
the stations). However, data from some stations were not useful as noted in
Table~\ref{tab:obs}. During the first observation, DE604 was using a wrong
observation table, whereas for the rest of the stations without valid data a
communication problem caused the data to be lost before getting to the
correlator in Groningen. Each 1-h observation contained twelve 4-minute target
scans, plus 1 minute between scans required for setup. For each target scan we
generated 30 beams to observe simultaneously 30 sources. The tile beam centre
was set to the source closest to the centre of the corresponding group, so the
30 sources are within an area of $\sim$2$^{\circ}$ from the pointing centre.
Each beam was allocated with 16 sub-bands with spanned bandwidth of
138.597--141.722 MHz, a frequency range chosen because it is near the peak of
the LOFAR sensitivity and is free of strong radio frequency interference (RFI).
Additionally, we observed a bright calibrator (3C380 or 3C295) for 5 minutes at
the beginning and at the end of each observation. The angular and temporal
separation of these scans from the target scans is too large for them to be
useful for the international stations, but they are used to calibrate the core
and remote stations, and we refer to it hereafter as the ``Dutch'' calibrator
source, to distinguish it from the discussion of primary and secondary
calibrators for the international stations. The separation between the Dutch
calibrator and the targets is  7.5--26$^{\circ}$ for 3C380, and 13--22$^{\circ}$
for 3C295, for observation 1 and 2, respectively.  We note that these
separations are predominantly North-South. For these Dutch calibrator scans a
single beam of 16 sub-bands was used, spanning the same bandwidth as the target
observations.

\begin{table}   
\caption{Log of the observations}
\label{tab:obs}
\centering
\begin{tabular}{@{\extracolsep{-3pt}}l c c}
\hline\hline
                        &  Observation 1 & Observation 2 \\ 
\hline
Date                    & 2013-05-02    & 2013-11-07    \\
UTC Time                & 06:00--07:15  & 04:20--05:30  \\
Dutch array calibrator  & 3C380         & 3C395         \\
Stations without valid data  & DE602, DE603  & DE603, SE607\tablefootmark{a}         \\
                             & DE604         &        \\
Failed Scans            & 2             & 1         \\
Sources scheduled       & 360           & 360            \\
Sources observed        & 300           & 330           \\
Mean elevation          & 80$^\circ$    & 55$^\circ$    \\
\hline
\end{tabular}
\tablefoot{
\tablefoottext{a}{Missing first half of the observation for SE607.}
}
\end{table}

\subsection{Target selection}

In our two observations, we applied different selection criteria in order to
cover a wide range of sources that could potentially be useful International
LOFAR primary calibrators. The selection was based on the WENSS catalogue
\citep{rengelink97}.  We used the peak flux density of the sources in this
catalogue, instead of the integrated flux density, because with a resolution of
54\arcsec any extended emission in WENSS will not contribute to the compact flux
at $\sim1\arcsec$ scales relevant to the LOFAR long-baseline calibration.  We
note that all WENSS peak flux densities in this paper include a correction
factor of 0.9 with respect to the original catalogue to place them to the RCB
scale \citep{roger73a}, as recommended by \cite{scaife12a}.  For the first
observation we selected an area of 11.6$^{\circ}$ radius centred on ($18^{\rm
h}30^{\rm m}$, +$65^{\circ}$), or Galactic coordinates $(l, b) = (94.84^{\circ},
+26.6^{\circ})$. The field contains 9251 sources from the WENSS catalogue
\citep{rengelink97} at 325~MHz.  We focused on the brightest sources by randomly
selecting 360 of the 1414 sources with a peak flux density above 180~mJy/beam
(at the WENSS resolution of 54\arcsec).  Ten of the selected sources are also
known cm~VLBI calibrators.  The second field observed was centred on ($15^{\rm
h}00^{\rm m}$, +$70^{\circ}$), or $(l, b) = (108.46^{\circ}, +43.4^{\circ})$,
with a radius of $4.86^{\circ}$.  Within this field, we selected any known
cm~VLBI calibrators with an integrated VLBI flux density above 100~mJy at 2.3
GHz; there were six such sources.  We completed our allocation of 360 sources by
selecting all sources with a WENSS peak flux density in the range 72--225
mJy/beam at 325 MHz.  In this way, we covered a representative sample of sources
with peak flux densities $>$72 mJy/beam at 325~MHz, as well as including a small
sample of sources which were known to be compact at cm wavelengths (the known
cm~VLBI calibrators). 

Due to a system failure part of the data were lost during the observations, and
two scans (60 sources) were missing during the first observation and one scan
(30 sources) during the second observation. Therefore, the actual number of
observed sources is 630, with 300 in the first field and 330 in the second.

\begin{figure*}[] 
\center
\resizebox{1.0\hsize}{!}{\includegraphics[angle=0]{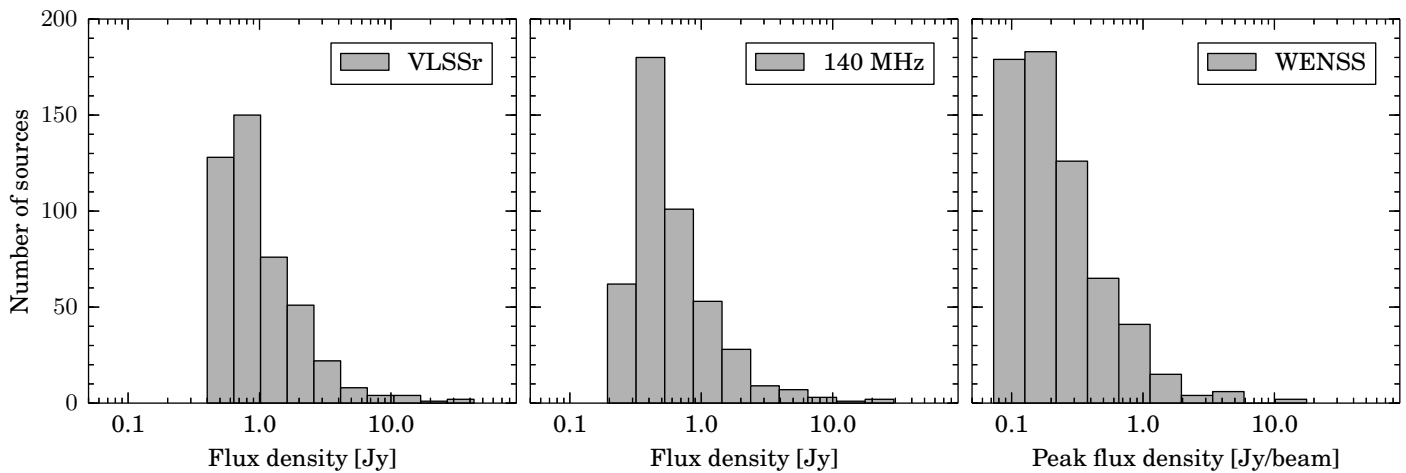}}
\caption{Flux density distribution of the observed sources in the VLSSr
catalogue at 74~MHz (\emph{left panel}, 447 sources), and WENSS at 325~MHz
(\emph{right panel}, 629 sources). The middle panel shows the estimated flux
density at 140~MHz interpolated from the two catalogues for sources with
counterpart on both catalogues.}
\label{fig:flux}
\end{figure*}

In Fig.~\ref{fig:flux} we show the distribution of flux density of the
counterparts of the observed sources found in the VLSSr catalogue at 74~MHz, 4~m
wavelength \citep{lane12a,lane14} and the WENSS catalogue at 325~MHz, 92 cm
wavelength.  Like the (corrected) WENSS flux densities, the VLSSr catalogue flux
scale is also set using the RCB flux density scale \citep{roger73a}.  Based on
these two values, we also show the distribution of the estimated flux density of
the sources at 140~MHz.

\subsection{Data reduction}\label{sec:datareduction}

The data reduction proceeded as follows. First, standard RFI flagging was
performed and the data averaged to a temporal resolution of 2 seconds and a
frequency resolution of 49~kHz (4 channels per LOFAR sub-band). Subsequently,
the complex gains of the core and remote stations were calibrated using standard
LOFAR tools (Black Board Selfcal [BBS], \citealp{pandey09a}) on the Dutch array
calibrators 3C380 and 3C295 using a simple point-source model. Beam calibration
was enabled, which corrects for the elevation and azimuthal dependence of the
station beam pattern in both linear polarisations before solving for complex
gain, using a model incorporating the station layout and up-to-date information
on station performance, such as failed tiles.  To derive the solutions we only
used baselines between core and remote stations, but not core--core or
remote--remote baselines. By not using the shortest baselines ($\lesssim3$~km)
we ensure that other nearby, bright sources do not contaminate the calibration
solutions, while by not using the longest baselines ($\gtrsim55$~km) the
point-source calibrator model remains valid.  Alternatively, all baselines can
be used if a detailed model of the structure of the source and all the nearby
sources are considered during the calibration.

A single amplitude and phase solution was derived for each sub-band, for each
scan on the Dutch array calibrator, and for each of the two observations.  We
verified that the phase solutions were relatively constant in time for the core
stations.  BBS does not allow interpolation of solutions, and so we only used
the solutions from a single scan (the final scan) to correct the core and remote
station gains for the entire dataset.   The gains of the international stations
were not included in this solution and are left at unity. The solution table was
exported with ``parmexportcal''\footnote{\label{note1}More info in:
http://www.astron.nl/radio-observatory/lofar/lofar-imaging-cookbook} in order to
be applied to data that were not observed simultaneously to the calibrator.
Finally, the solutions were applied to all sources using
``calibrate-stand-alone'' with a blank model as input.  Note that this scheme
applies the {\em a priori} station beam corrections to all stations (including
the international stations); the additional `solved' corrections are present
only for the core and remote stations, but not for international stations.

With the core stations now calibrated, it was possible to form a coherent ``tied
station'', hereafter TS001, from all of the core stations. Because the beams are
already centred directly on individual sources, the phasing-up of the core
stations does not require additional shifts, which significantly reduces the
processing time. TS001 is formed by summing baseline visibilities with the NDPPP
task ``StationAdder''.  The New Default Pre-Processing Pipeline (NDPPP) forms a
major component of the standard LOFAR imaging pipeline, and is described in
\citet{heald10a} and, with the most up-to-date information, in the LOFAR imaging
cookbook (see footnote \ref{note1} in page \pageref{note1}).  After this step,
all original visibilities with baselines to core stations were discarded using
the NDPPP task ``Filter'' to reduce data volume.

Subsequently, for each source we combined the 16 sub-bands together into a
single measurement set using NDPPP.  To avoid the rapid phase changes with
frequency introduced into linear polarisation data on long baselines by
differential Faraday rotation, we converted the data to circular polarisation
using the Table Query Language (TAQL) to operate on the measurement set data
directly\footnote{A measurement set can be converted to circular polarisation
    using: update <filename.ms> set DATA = mscal.stokes(DATA,'circ') and update
<filename.ms>/POLARIZATION set CORR\_TYPE=[5,6,7,8]}.  As the effects of the
station beam were already calibrated, this is a simple operation. However, we
note that since BBS calibrates the XX and YY components independently, an
overall phase-offset between X and Y may remain.  Although this could not be
corrected for this project, we checked that the residual RL/LR leakage is a
minor contribution and not critical for this kind of detection experiment.
Finally, we converted the measurement set to UVFITS format in order to proceed
with the phase calibration using the AIPS software package \citep{greisen03a}.
At this stage, the data volume had been reduced from the original $>$4000~GB
from a one hour observation to 35~GB, and a much more sensitive tied-array
station had been generated to aid the derivation of calibration solutions to the
international stations.

In AIPS we conducted a phase calibration by fitting the station-based phases,
non-dispersive delays, and rates using the task FRING. We searched solutions
using all international stations and the combined core station, TS001, which was
used as a reference station.  No models were provided for any sources, and hence
all sources were assumed to be point-like. The maximum search window for delay
and rate was set to 1000~ns and 50~mHz, respectively. These windows were
motivated by the largest values expected from ionospheric effects. However, a
1000~ns delay search (or 50~mHz rate search) corresponds to the effect incurred
by a source up to 5 arcminutes (13 arcminutes) away from the nominal position,
for the shortest international station to LOFAR core baselines.  Confusing
sources more than 5 arcminutes away from the nominal source direction are
therefore filtered out. Additionally, the tied station has a synthesised field
of view of about 3 arcminutes, and thus sources further away than this distance
will contribute less to the fit.  The solution interval was set to the scan
duration (4 minutes), and so only one solution per polarisation was derived per
station, per source.  We extracted the fit solutions from AIPS for further
analysis using the \verb+ParselTongue+ software package \citep{kettenis06a}.

\section{Analysis}\label{sec:analysis}

With these snapshot observations it is not possible to conduct an amplitude
calibration of the international stations.  None of the sources observed have a
model that could reasonably be extrapolated to our observing frequency and
resolution, and so self-calibration is not feasible.  At present, the
instrumental gains within LOFAR are not tracked with time\footnote{This is
planned to change with the new ``COBALT'' correlator recently commissioned.},
and so making a sufficiently accurate {\em a priori} calibration is also not
feasible.  With a longer observation (as would be typical for a normal science
observation), it would be possible to bootstrap from approximate amplitude
corrections for the international stations and image/self-calibrate the target
source, but the $uv$ coverage in our snapshot observations is too sparse for
such an approach.

\begin{figure}[] 
\center
\resizebox{1.0\hsize}{!}{\includegraphics[angle=0]{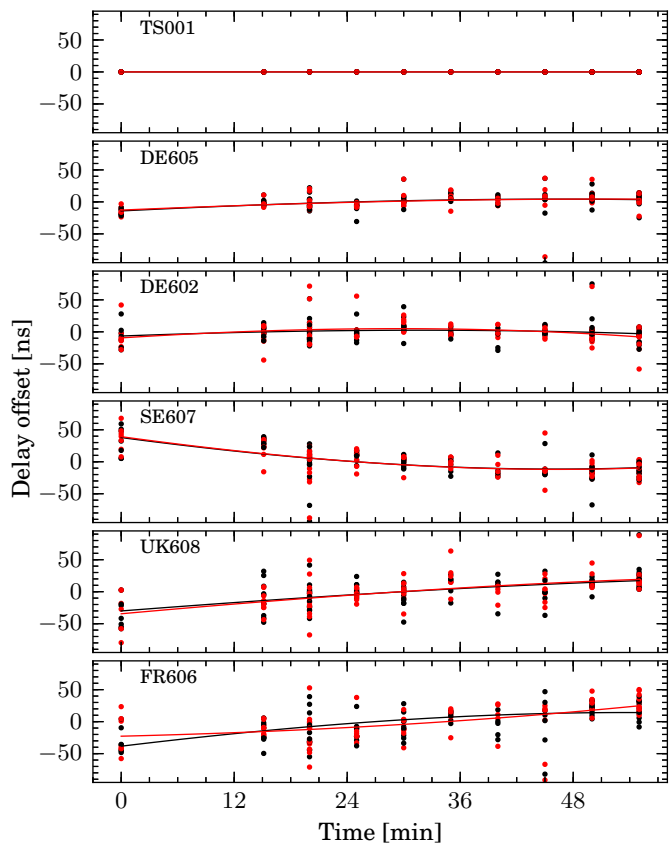}}
\caption{Station delay offsets as a function of time for those sources with
fringe solutions for all stations. Up to 30 sources are observed
simultaneously at each time interval. The delays are referenced to the
average for each station, quoted in Table.~\ref{tab:delays}. Black circles
correspond to left-hand polarisation delays, $\tau_{\rm L}$, and red circles
correspond to right-hand polarisation delays, $\tau_{\rm R}$. The formal
uncertainties from the fringe fit are smaller than the markers.}
\label{fig:drift1}
\end{figure}

\begin{figure}[] 
\center
\resizebox{1.0\hsize}{!}{\includegraphics[angle=0]{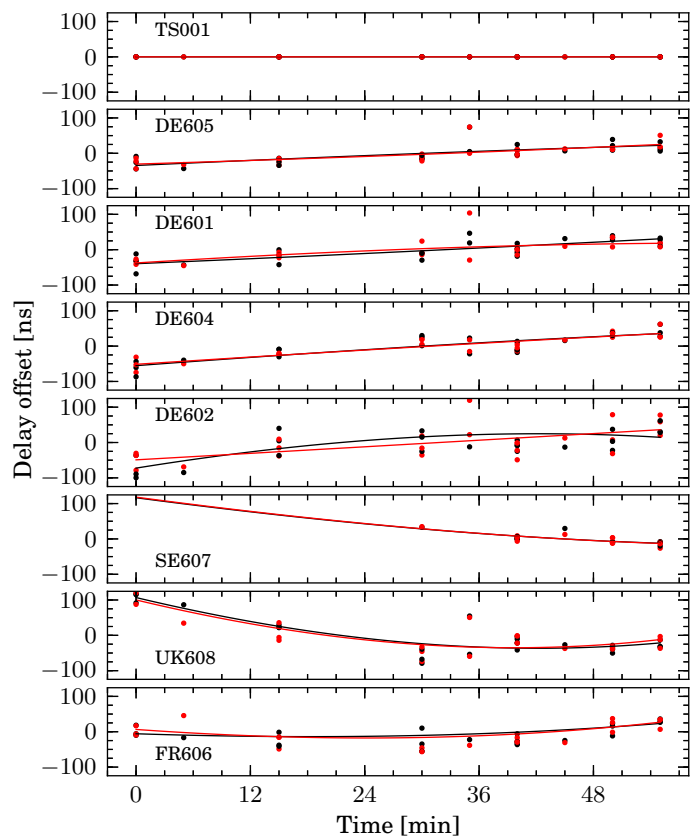}}
\caption{Same as Fig.~\ref{fig:drift1} but for the second observation.}
\label{fig:drift2}
\end{figure}

Instead, we base our analysis of the compactness of the sources on the phase
information of the data, in particular on the capability of each source to
provide good delay/rate solutions for the international stations. In order to
identify ``good'' solutions, we first need to identify an approximation to the
true station delay.  In order to do this, we selected only the sources that
provided delay/rate solutions with a signal-to-noise ratio above 6 for all
stations. In Figs.~\ref{fig:drift1} and \ref{fig:drift2} we show the evolution
of the delay with respect to time. We plot the delay offset with respect to the
average value for each station, which is quoted in Table~\ref{tab:delays}.  We
fitted a polynomial of degree 3 to the delay evolution with time and plotted it
using a solid line. We measured the delay rate as the average delay derivative
with time for each station. In summary, in Table~\ref{tab:delays} we quote the
average delay per station and polarisation, which is the reference for the
offsets in Figs.~\ref{fig:drift1} and \ref{fig:drift2}, their uncertainties,
computed as the standard deviation from the fitted polynomial, and the measured
delay rate, computed as the average delay derivative for each station.

\begin{table}   
\caption{Average station-based delay solutions in left ($\tau_{L}$) and right ($\tau_{L}$) circular polarisation, fitted to the subset of sources which gave good solutions on all stations.}
\label{tab:delays}
\centering
\begin{tabular}{l r@{ $\pm$ }l r@{ $\pm$ }l r}
\hline\hline
Station & \multicolumn{2}{c}{$\tau_{\rm L, 0}$} &  \multicolumn{2}{c}{$\tau_{\rm R, 0}$} & $\partial\tau /\partial t$  \\
       & \multicolumn{2}{c}{ [ns]}  &  \multicolumn{2}{c}{ [ns]}  &
[ns~h$^{-1}]$  \\
\hline
   &  \multicolumn{5}{c}{Observation 1}   \\
\hline
TS001    &    0 &    0 &     0 &    0 &    0   \\
DE605    &  104 &   14 &   106 &   13 &   19   \\
DE602    &  184 &   15 &   185 &   18 &    3   \\
SE607    &  109 &   26 &   106 &   26 &  -52   \\
UK608    &  100 &   20 &   103 &   20 &   55   \\
FR606    &   59 &   44 &    56 &   33 &   55   \\
\hline
   &  \multicolumn{5}{c}{Observation 2}   \\
\hline
TS001    &    0 &    0 &     0 &    0 &    0   \\
DE605    &  126 &   19 &   126 &   18 &   61   \\
DE601    &  -11 &   18 &    -9 &   25 &   69   \\
DE604    &  241 &   19 &   229 &   15 &   97   \\
DE602    &  377 &   48 &   366 &   38 &   94   \\
SE607    &    0 &   44 &     0 &   44 & -142   \\
UK608    & -117 &   27 &  -110 &   29 & -131   \\
FR606    &  102 &   49 &   106 &   54 &   27   \\
\hline
\end{tabular}
\end{table}

If the model applied at correlation time were perfect, all stations would see a
delay offset of zero for all sources, but deviations are produced by several
factors. Table~\ref{tab:expecteddelay} summarises the main contributions and the
time scale in which they change. First, errors in station positions (and
currently in a much lower level errors in the the Earth orientation parameters,
EOPs) used by the correlator produce variability of about $\pm75$~ns with a 24~h
periodicity.  The current correlator model used by LOFAR is insufficiently
accurate, and this source of error can be expected to be greatly reduced in the
near future.  Instabilities in the rubidium clocks can produce delay rates up to
20 ns per 20 min, which corresponds to about a radian per minute at 150~MHz
\citep{van-haarlem13a}.  In total, non-dispersive instrumental delays of up to
$\sim100$~ns and delay rates of up to $\sim$20~ns~h$^{-1}$ are expected.
Second, for any given source, errors in the {\em a priori} centroid position
(which is based on the WENSS position, with a typical error of 1.5\arcsec)
and/or extended structure on subarcsecond scales contribute an additional delay
offset.  The maximum baseline between an international station and the LOFAR
core is 700 km (for the FR606 station in France); a positional error of
1.5\arcsec\ will lead to a delay error of $\sim$15 ns on this baseline.

\begin{table}   
\caption{Approximate delay contributions at 140~MHz to a 700~km baseline.}
\label{tab:expecteddelay}
\centering
\begin{tabular}{l r r}
\hline\hline
Effect  & Delay  & Time scale \\
\hline
     \multicolumn{3}{c}{Non-Dispersive}   \\
\hline
Correlator model error      & $\sim75$~ns         &  24h (periodic)   \\
Station clocks              & $\sim20$~ns        & $\sim$20~min \\
Source position offset (1.5\arcsec)        & $\sim15$~ns        & --  \\
\hline
     \multicolumn{3}{c}{Dispersive}   \\
\hline
Slowly varying ionosphere   & $\sim300$~ns      & $\sim$hours \\
Rapidly varying ionosphere  & $\ga$10~ns         & $\sim10$~min  \\
Differential ionosphere     & 5~ns/deg sep.      & -- \\
(source elevation 60 deg)   &                    &  \\

\hline
\end{tabular}
\end{table}

The ionospheric contribution to the delay changes as a function of time,
position, and zenith angle.  The magnitude of the changes depend on the Total
Electron Content (TEC) of the ionosphere, with a delay of $\tau_{\rm
ion}=c^{2}r_{\rm e}/(2\pi\nu^{2})\times {\rm TEC}$, being $c$ the speed of
light, $r_{\rm e}$ the classical electron radius, and $\nu$ the observed
frequency, and TEC is usually measured in TEC Units (1${\rm
TECU}=10^{16}$~electrons~m$^{-2}$).  The TEC can can be estimated using models
derived from observations of GPS satellites.  Models are available from
different institutes, such as the Jet Propulsion Laboratory (JPL), the Center
for Orbit Determination in Europe (CODE), the ESOC Ionosphere Monitoring
Facility (ESA), among others. We used the models produced by the Royal
Observatory of Belgium GNSS group\footnote{http://gnss.be/}, which are focused
on Europe and have an angular resolution of 0.5 degrees and a temporal
resolution of 15 minutes. The models contain information on the vertical total
electron content (VTEC) during the two observations, which were conducted
shortly after sunrise and at night, respectively. We note that the TEC values
above the stations are a lower limit of the slant ionospheric contribution that
depends on the source elevation at each station. More details can be found in,
for instance, \cite{nigl07} and \cite{sotomayor13a}.

The VTEC above the international stations was about 12--16 and 4--8 TECU for
observation 1 and 2, respectively.  These values correspond to an ionospheric
delay at 140~MHz of about 850--1100~ns, and 300--540~ns, respectively. These
values were approximately constant for observation 2, and were changing at a
rate of 90--120~ns~h$^{-1}$ for observation 1. These changes are expected, based
on $\sim$0.1--0.2~TECU variations in 10 minutes seen with the VLA at 74~MHz by
\cite{dymond11}, which corresponds to about 10~ns at 140~MHz. Although all VTEC
follow a similar 24-h trend strongly correlated with the Sun elevation, the
short-term (10--60 minute) variations between the widely separated international
stations are virtually uncorrelated.  The ionospheric contribution typically
dominates the total delay and delay rate for international LOFAR stations.
However, for sources observed simultaneously, up to 30 per scan, the ionospheric
contribution should be similar because they are separated by 4$^{\circ}$ at
most. We have used VLBI observations (VLBA project code BD152) at 300~MHz, or
1~m wavelength, of bright and compact pulsars at different angular separations
to obtain a rough estimate of the delay difference between sources separated
1--5~degrees at elevations of 50--80$^{\circ}$. As a first approximation we
estimated that the dispersive delay difference between sources at different
lines of sight should be about 5~ns per degree of separation, for a source
elevation of 60$^{\circ}$.

Noise is the final contribution to the delay offset, and depends on the
brightness of the source and the sensitivity of the station.  The dispersions
shown in Figs.~\ref{fig:drift1} and \ref{fig:drift2} are due to source position
and structure errors, differential ionosphere, as well as noise. As shown above,
errors of up to several tens of ns can be expected for any individual source
from both source position errors and differential ionosphere for our observing
setup.  In this analysis we assume that this contribution is random for any
given source, and that the delays measured at each time should cluster around
the real instrumental + mean ionospheric delay for each station at the time of
the scan.

\subsection{Quality factor}
To complete our analysis we compute a discrete quality factor, $q$, for each
source, assigning $q=3$ to bright and compact sources (i.e.  good primary
calibrators), $q=2$ to partially resolved sources, and $q=1$ to resolved or
faint sources. The quality factor $q$ is based on how many international
stations can be fringe fitted using a particular source to give a satisfactory
station delay. A source produces a good delay solution if the fit has a S/N
above 6, the difference between right- and left-circular polarisations is below
30~ns, and the deviation from the average delay (see Table~\ref{tab:delays}) is
below 300~ns. For each source, the factor $q$ is assigned depending on the
number of satisfactory delay solutions found. For an observation with N
international stations, $q=3$ is assigned if the number of satisfactory delay
solutions is $\geq N-1$; i.e., at most one station failed to provide a solution.
Failure of only one station is not uncommon on these observations with only one
short scan. $q=3$ sources are almost certainly suitable primary calibrators. 

A quality factor $q=1$ corresponds to sources with very low number of good
calibrated stations, where the number of failed solutions exceeds 3.  These
sources are heavily resolved on International LOFAR baselines and are almost
certainly unsatisfactory primary calibrators.  The intermediate category $q=2$
corresponds to sources where a significant number of stations see good
solutions, but at least two fail. This group would consist primarily of sources
with significant structure on arcsecond scales.  Some of these sources may be
suitable for calibration if a good model of the source structure could be
derived, but many would simply contain insufficient flux density on subarcsecond
angular scales.  The total number of sources in each group is listed in
Table~\ref{tab:quality}.

\begin{table}   
\caption{Number of sources as a function of quality factor. $q=3$ corresponds to suitable primary calibrators.}
\label{tab:quality}
\centering
\begin{tabular}{l c c c c}
\hline\hline
        &  $q=1$ & $q=2$ & $q=3$ & Total    \\ 
\hline
Observation 1  & 144  &  86 & 70 & 300    \\
Observation 2  & 234  &  80 & 16 & 330    \\
Total          & 378  & 166 & 86 & 630    \\
               & 60\%  & 26\% & 14\% & 100\%   \\
\hline
\end{tabular}
\end{table}
%

A catalogue containing the list of sources, basic information in the WENSS,
VLSSr, and NVSS catalogues, and the quality factor $q$ obtained here will be
available in a table online. A sample of some of the columns and rows of the
catalogue is shown in Table~\ref{tab:catalog}.

\begin{table*}  
\caption{Catalogue of observed sources showing the name in the WENSS (or VLBI if
available) catalogue, J2000 position, peak flux density or integral flux density
in the considered catalogues, and quality factor computed in this paper.}
\label{tab:catalog}
\centering
\begin{tabular}{l ll r@{ $\pm$ }l r@{ $\pm$ }l c c}
\hline\hline
 Name & Right Ascension & Declination & \multicolumn{2}{c}{$S_{\rm peak,
 WENSS}$} & \multicolumn{2}{c}{$S_{\rm NVSS}$} & $S_{\rm VLSSr}$ &  $q$ \\
      & [hh:mm:ss]      & [dd:mm:ss]  & \multicolumn{2}{c}{[mJy~beam$^{-1}$]} & \multicolumn{2}{c}{[mJy]}  & [Jy]  &   \\
\hline
WNB1927.8+7119  & 19:27:22.13 & 71:25:41.7 & 276 & 5  & 86 & 3 & 1.1 & 2 \\
J1927+7358      & 19:27:48.06 & 73:58:01.7 & 4165 & 5 & 3900 & 120 & 8.1 & 3 \\
WNB1928.8+7032  & 19:28:30.34 & 70:38:37.9 & 987 & 5  & 312  & 11 & 3.5 & 1 \\
WNB1930.8+7121  & 19:30:20.79 & 71:27:34.8 & 252 & 5  & 72   & 2 & 1.2 & 2 \\
WNB1931.9+7203  & 19:31:18.77 & 72:10:22.7 & 921 & 5  & 261  & 8 & 2.5 & 2 \\
WNB1935.7+7338  & 19:34:47.04 & 73:45:13.7 & 515 & 5  & 166  & 6 & 1.7 & 2 \\
WNB1937.1+7056  & 19:36:46.90 & 71:03:23.5 & 194 & 5  & 50   & 2 & 0.6 & 2 \\
WNB1937.3+7127  & 19:36:51.11 & 71:34:47.1 & 359 & 5  & 125  & 4 & 1.2 & 1 \\
WNB1939.2+7235B & 19:38:40.05 & 72:42:57.5 & 353 & 5  & 98   & 3 & 1.1 & 2 \\
WNB1941.7+7053  & 19:41:23.19 & 71:00:48.1 & 1017 & 5 & 359  & 11 & 2.0 & 3 \\
WNB1945.9+7240  & 19:45:16.63 & 72:47:57.2 & 1956 & 5 & 810  & 30 & 4.1 & 3 \\
WNB1950.8+7213  & 19:50:19.71 & 72:21:37.6 & 210 & 5  & 71   & 3 & -- & 2 \\
WNB1954.4+7039  & 19:54:10.86 & 70:47:28.1 & 225 & 4  & 64   & 2 & 0.7 & 1 \\
\hline
\end{tabular}
\end{table*}

\section{Results}\label{results}

\subsection{Sky distribution} 

\begin{figure}[t]    
\center
\resizebox{1.0\hsize}{!}{\includegraphics[angle=0]{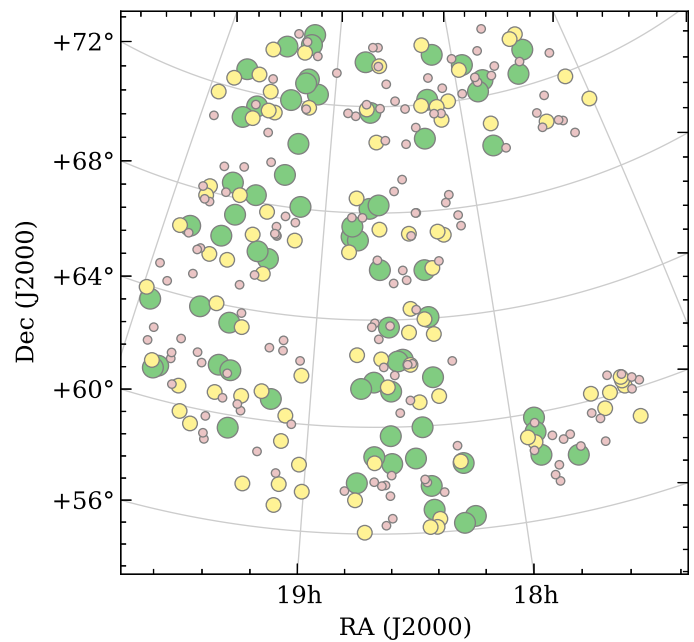}}
\caption{Sky distribution of the sources in observation 1, with markers
indicating good primary calibrators, $q=3$ (big green circles), potentially good
primary calibrators, $q=2$ (medium-size yellow circles), and resolved and/or
faint sources, $q=1$ (small red circles).  The gap in right ascension
17--18$^{\rm h}$ and declination 60--68$^{\circ}$ is caused by the failure of
two scans, as described in the text.}
\label{fig:sky1}
\end{figure}

\begin{figure}[t]    
\center
\resizebox{1.0\hsize}{!}{\includegraphics[angle=0]{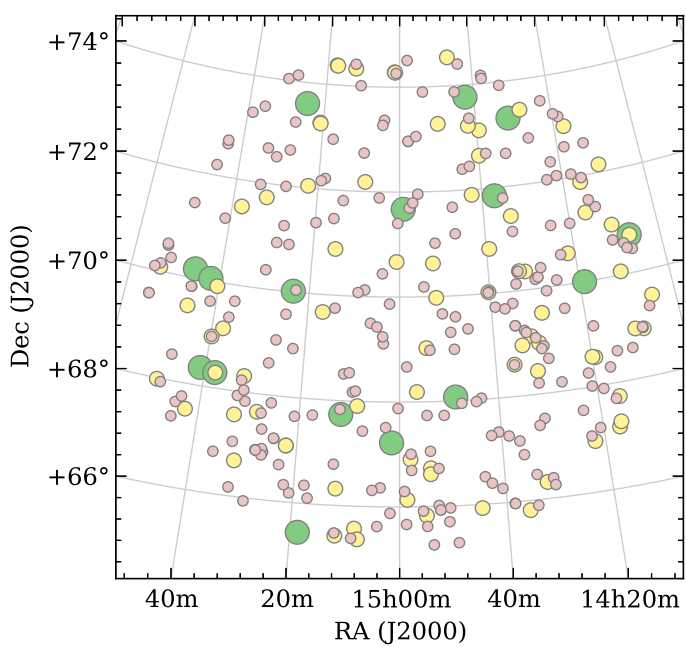}}
\caption{Sky distribution of the sources in observation 2, with markers
indicating good primary calibrators, $q=3$ (big green circles), potentially good
primary calibrators, $q=2$ (medium-size yellow circles), and resolved and/or
faint sources, $q=1$ (small red circles)).}
\label{fig:sky2}
\end{figure}

With the selection of good ($q=3$) and potentially good ($q=2$) primary
calibrators for the LOFAR long baselines provided by the international stations
we can study the properties of this source population. In Figs.~\ref{fig:sky1}
and \ref{fig:sky2} we show the distribution of observed sources for observation
1 and 2, respectively, and the corresponding quality factor indicated by the
size/colour of the markers. The gap in Fig.~\ref{fig:sky1} in right ascension
17--18$^{\rm h}$ and declination 60--68$^{\circ}$ is due to the loss of 2 scans
(60 sources).  In the second observation the sources were distributed in three
passes through four different sectors, and thus the loss of one scan only
produced a lower number of sources in the north-western sector. The distribution
of good sources does not depend on the distance to the Dutch array calibrator,
used to phase-up the core stations.  Also, no significant bias is seen with
respect to right ascension or declination. Therefore, the distribution of likely
good primary calibrators ($q=3$) is uniform in these two fields.

\subsection{Flux density, spectral index and extended emission}

We used the information in the VLSSr, WENSS and the NVSS \citep{condon98}
surveys to study the correlation of flux density, spectral index, and
compactness (as seen from low angular resolution data) with their suitability as
a primary calibrator, as evidenced by the quality $q$.

To compute the spectral index of the sources we have used the integrated flux
density in VLSSr and NVSS, and the peak flux density in WENSS. The three surveys
all have different resolutions (75, 45, and 55 arcseconds, respectively), which
makes a direct comparison of flux density and hence spectral index difficult
(regardless of whether peak or integrated flux density is used). Fortunately,
relatively few sources are resolved in these surveys, and so for each survey we
chose to simply use the primary value given in the survey catalog in question.
The small biases which are introduced to the spectral indices following measured
analysis.

\begin{figure}[]    
\center
\resizebox{1.0\hsize}{!}{\includegraphics[angle=0]{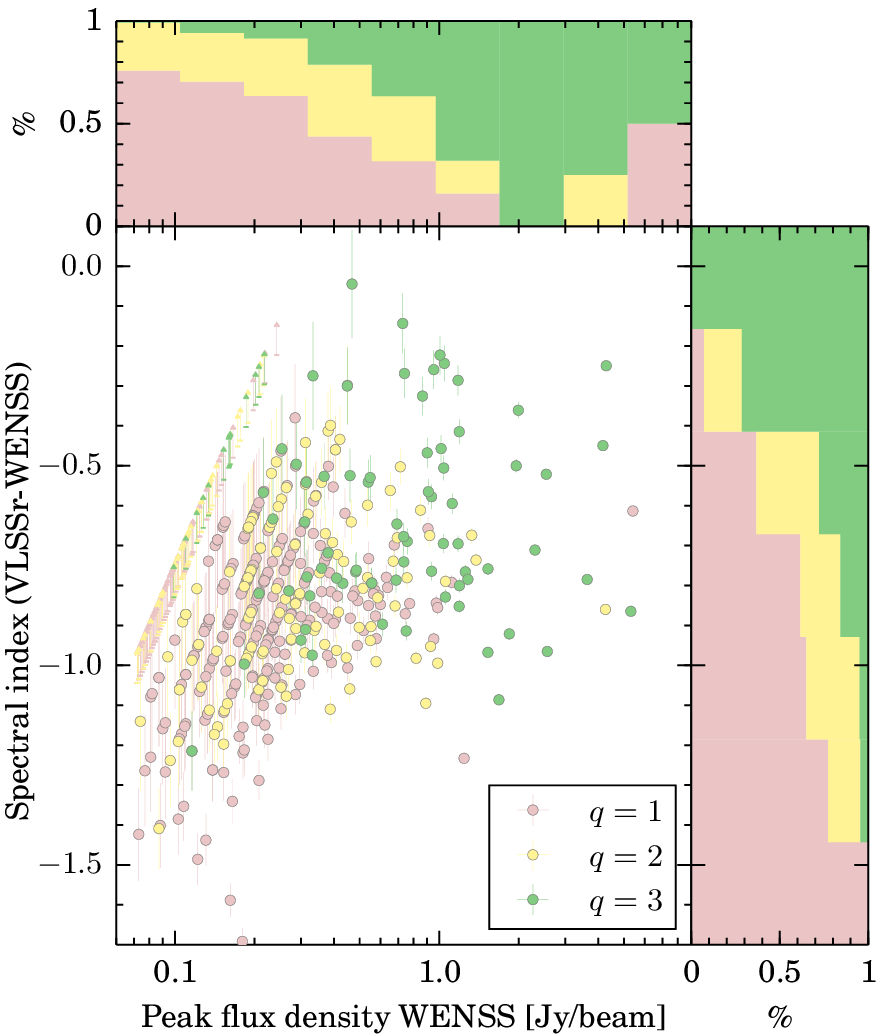}}
\caption{Source quality as a function of low-frequency spectral index and WENSS
peak flux density. Red, yellow, and green colours represent $q=1, 2$, and 3,
respectively.  $q=3$ corresponds to suitable primary calibrators. The histograms
show the percentage of sources at each quality value. The histogram on the right
does not include the sources with upper limits in their spectral indexes. 
The uncertainties of some of the values are smaller than the symbols, specially
for the WENSS peak flux density. The stripes at low peak flux densities are due
to sources detected at multiples of the rms noise of VLSSr.}
\label{fig:sp_wenss}
\end{figure}

In Fig.~\ref{fig:sp_wenss}, we show the distribution of sources as a function of
low frequency spectral index, measured with VLSSr and WENSS, and the peak flux
density in WENSS.  The histograms show the percentage of sources with each $q$
factor for different ranges of these two variables.  We see that brighter
sources are more likely to be good primary calibrators, which is unsurprising:
for a faint source to be a suitable calibrator, almost all of the flux density
must be contained in a subarcsecond component, whereas a bright source can
possess significant extended emission and still contain sufficient flux density
in a compact component.  Sources which are brighter than 1~Jy/beam at 325~MHz
are more likely than not to be a satisfactory primary calibrator, whereas
sources of 0.1~Jy/beam are extremely unlikely to be suitable.

Fig.~\ref{fig:sp_wenss} also shows that sources with a flatter low-frequency
spectrum (measured in this instance from 74 to 325 MHz) are much more likely to
be satisfactory primary calibrators.  Again, this is unsurprising:
steep-spectrum emission is typically associated with extended radio lobes, which
would be resolved out with our subarcsecond resolution.  Sources with a
low-frequency spectral index $> -0.4$ (where $ S \propto \nu^{\alpha} $) are
almost always suitable primary calibrators.

\begin{figure*}[]   
\center
\resizebox{1.0\hsize}{!}{\includegraphics[angle=0]{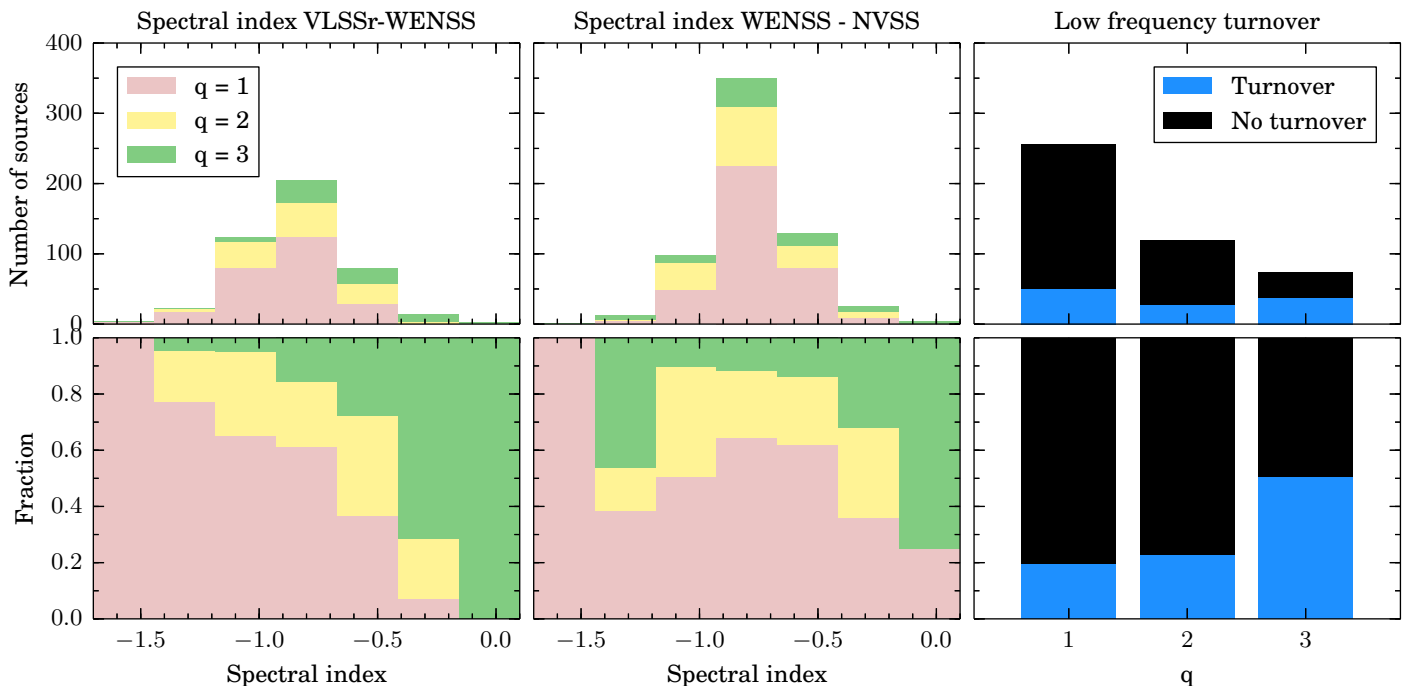}}
\caption{Source quality as a function of low-frequency spectral index (left
panel) and high-frequency spectral index (middle panel), with absolute number
of sources and percentage at each quality factor. The lower number of sources
with low-frequency spectral index is because fewer sources have a VLSSr
counterpart. In the right panel we show the number of sources with a given
quality showing or not showing a low frequency turnover.}
\label{fig:spindex}
\end{figure*}

However, the spectral index at higher frequencies (computed in this instance
from 325 to 1400 MHz, using WENSS and NVSS) is a much poorer predictor of
calibrator suitability.  Fig.~\ref{fig:spindex} shows the distribution of
quality factor with low and high frequency spectral index. The difference in
predictive power is obvious: by selecting a source with spectral index $>-0.6$
the chance probability of the source having $q=3$ is 51\% if we use the
low-frequency spectral index and only 36\% if we use the high-frequency spectral
index. The right panel of Fig.~\ref{fig:spindex} shows the relative prevalence
of a spectral turnover (where the low frequency spectral index is flatter than
the high frequency spectral index) for the three quality bins.  Good primary
calibrator sources ($q=3$) are more likely to see a spectral turnover.

Finally, for each source, we can check whether it was resolved in the VLSSr and
NVSS catalogues. In Table~\ref{tab:compact} we show the number of sources listed
as resolved and unresolved in each catalogue and the associated quality factor
percentages.  The low angular resolution of VLSSr means that almost all sources
in the catalogue are unresolved and few conclusions can be drawn.  On the other
hand, one third of the sources in NVSS are resolved by that survey, and we can
determine significant trends.  If a source is resolved by NVSS, it is very
unlikely to be a good primary calibrator (more than 5 times less likely than if
it is unresolved in NVSS). Table~\ref{tab:compact} also shows the percentage of
cm-VLBI calibrators that are compact in our data. 6 out of the 15 cm-VLBI
calibrators are not good calibrators at 140~MHz, 3 with $q=1$ and 3 with $q=2$.
Four of them have inverted or gigahertz peaked spectra and are probably too
faint at 140 MHz. The remaining two VLBI sources that proved to be
unsatisfactory calibrators, J1722+5856, and J1825+5753, have a flat-spectrum
VLBI core with moderate flux density ($\sim$150--200~mJy); they may have
decreased in flux density since the VLBI observations, or possibly the core
exhibits a low-frequency turnover. Based on our small sample of cm-VLBI
calibrators, compactness at cm wavelengths is also a good predictor of
suitability as an International LOFAR primary calibrator. A sufficiently bright
cm-VLBI calibrator, accounting for spectral index, will have enough compact flux
at 140~MHz with very high reliability. Although the correlation between being a
cm-VLBI calibrator and being a good LOFAR calibrator is clear from these data,
we note that this conclusion relies on a very low number of sources (15) and
better statistics are needed to derive more accurate statistics.

\begin{table}   
\caption{Total number and percentage of sources as a function of quality factor
and source compactness.}
\label{tab:compact}
\centering
\begin{tabular}{l r c c c }
\hline\hline
                &  Total      & \multicolumn{3}{c}{\% of sources}    \\ 
\cline{3-5}
                &    & $q=1$ & $q=2$ & $q=3$    \\ 
\hline
VLSSr compact    &    407 & 58 & 27 & 15  \\
VLSSr resolved   &     39 & 51 & 23 & 26  \\
\hline
NVSS compact    &    401 & 49 & 31 & 20  \\
NVSS resolved   &    223 & 79 & 17 & ~4  \\
\hline
VLBI calibrator &    15 & 20 & 20 & 60  \\
\hline
\end{tabular}
\end{table}

\subsection{Calibrator selection strategies and sky density}\label{sec:skydensity}

\begin{figure*}[ht]   
\center
\resizebox{1.0\hsize}{!}{\includegraphics[angle=0]{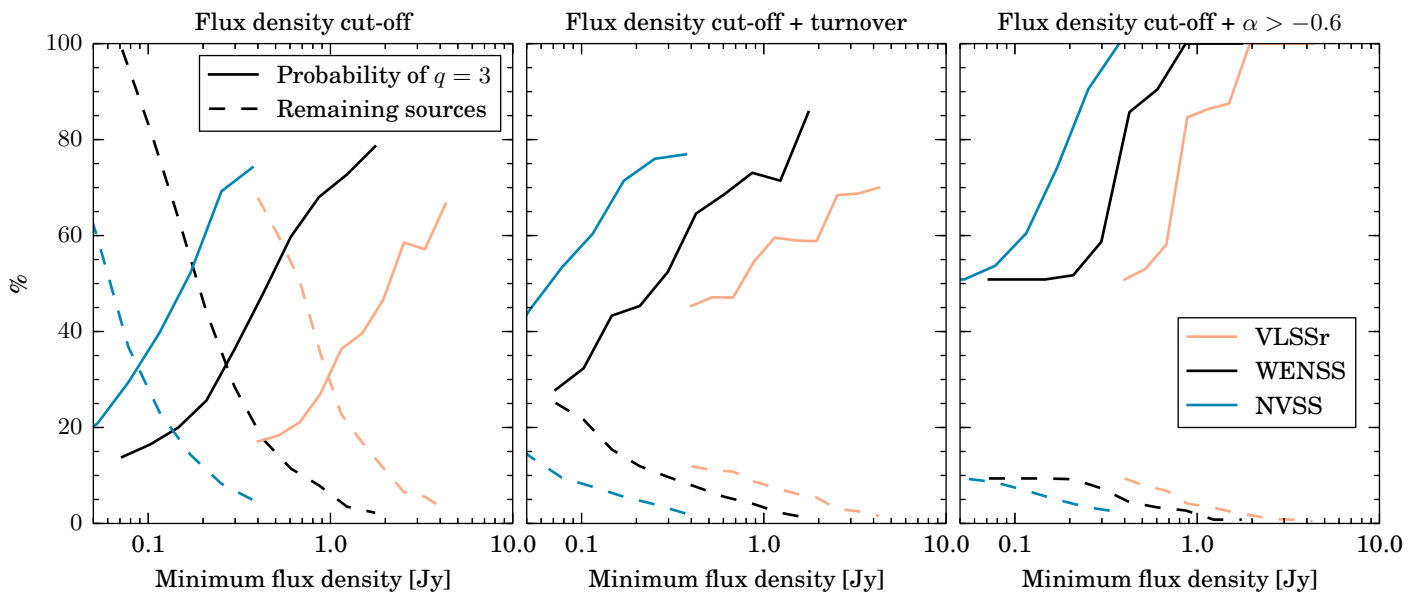}}
\caption{The effect of applying different preselection criteria to improve the
detection fraction of calibrator sources, including a lower limit on flux
density (left panel), lower limit on flux density plus requiring a low frequency
spectral turnover (middle panel), or lower limit on flux density plus a lower
limit to spectral index as calculated between VLSSr and WENSS (right panel).  The
3 colours correspond to imposing the lower flux density limit on the value
obtained from VLSSr (orange), WENSS (black) or  NVSS (blue).  In each panel, the
dashed line shows the fraction of the total sample which remains as the lower
limit to flux density is raised, while the solid line shows the fraction of that
remaining sample which are good calibrators. Imposing a spectral turnover or low
frequency spectral index requirement can reduce the sample size by a factor of
10 whilst still discovering almost half of the total acceptable calibrators.}
\label{fig:flux_cutoff}
\end{figure*}

We have shown above that peak flux density, low-frequency spectral index, and
compactness on scales of tens of arcseconds are all good predictors of primary
calibrator suitability for LOFAR.  To help selecting a sample of potential
calibrators we plotted in Fig.~\ref{fig:flux_cutoff} the percentage of good
primary calibrators, with $q=3$, as a function of the minimum peak flux density
imposed to the sample, for three different selection criteria.  For example, the
left panel shows that we expect 20\% of the sources with WENSS peak flux density
above 0.2~Jy/beam to be good primary calibrators. If additionally we impose that
the sources have a low-frequency turnover (middle panel) the probability of
having a good source increases to 45\%, whereas selecting sources with flat
low-frequency spectrum (right panel) increases the chances to 50\%. However, a
restrictive criterion comes with a reduction of the number of sources in the
sample, as shown by the dashed lines.  The low number of sources in the right
panel compared to the middle panel is also because many of the faint sources are
not detected with VLSSr, and thus a measurement of the spectral index is not
available, although it is still possible to infer that there is a turnover when
the source is not detected.

However, rigorous preselection is unlikely to be necessary in practice.  Based
on these results, we can extrapolate to the density of suitable primary
calibrators on the sky.  The field observed in the first epoch contains 1200
WENSS sources with peak flux density above 180~mJy/beam, from which we observed
300 and found 70 good primary calibrators, giving an estimate of 280 good
calibrators in the effective 350 square degrees observed. Therefore, the density
of good primary calibrators for the criteria of epoch 1 is 0.8 per square
degree. The 16 good primary calibrators out of 330 sources with WENSS peak flux
densities between 72 and 225~mJy/beam found in the 62 square degrees of
observation 2 provide 0.24 good primary calibrators per square degree.  To
obtain the density of the whole sample we corrected for the sources being
counted twice in the same flux density range: WENSS peak flux densities between
180 and 225~mJy/beam. We conclude that the density of good primary calibrators
with 325~MHz peak flux density $>72$~mJy/beam is approximately 1.0 per square
degree.  Unfortunately, the low statistics of the overlap region, with 3 and 4
good primary calibrators, respectively, prevents us from obtaining a significant
uncertainty on this density.

From the WENSS survey, there are $\sim7.6$ sources above 72~mJy/beam per square
degree, 14\% of them expected to be good primary calibrators. After an overhead
of 4 minutes for the calibration of the LOFAR core stations, we can survey 30
sources per 4 additional minutes. That means an area covering 3 square degrees
(a radius of $\sim$1$^{\circ}$) around a target source can be inspected for
primary calibrators in just 10 minutes. Without any other preselection, the
likelihood of identifying at least one usable calibrator among 30 WENSS sources
is 98.9\%.  Depending on the specific requirements of a project and the
characteristics of the field around the target source, this probability can be
increased by observing 60 sources up to 1.6$^{\circ}$ around the target in
15~minutes, or by setting additional selection criteria (see
Fig.~\ref{fig:flux_cutoff} or Table~\ref{tab:compact}).  Such a calibrator
search could easily be undertaken in the weeks prior to a science observation.

Once a primary calibrator has been identified, a secondary phase calibrator
closer to the target could be identified if the target itself will not be strong
enough for self-calibration. This is more efficiently conducted in a separate,
second observation, because the full bandwidth would be required to search for
fainter sources.  This could be set up in an identical manner to a typical
International LOFAR science observation, with the pointing centre set to be
midway between the primary calibrator and the target field.  After correlation,
the full-resolution visibility dataset can be shifted and averaged multiple
times, to the position of the primary calibrator and to the position of all
candidate secondary calibrator sources.  Since 30$\times$ more bandwidth is
used, again a 10 minute observation would suffice to identify useful secondary
calibrators (those with a peak flux density $\gtrsim$5 mJy/beam).  

As an alternative to a separate, short observation before the science
observation, a small subset of the data from the science observation itself
could be used to search for a secondary calibrator.  The advantage of a short
search in advance is that the secondary calibrator-target separation is known,
which could inform the selection of observing conditions (if a good calibrator
is present, poorer quality ionospheric conditions could be tolerated, for
instance).

Additionally, the same approach used for finding and using secondary calibrators
can be applied several times towards different sky directions to survey the
whole station beam. Relatively faint secondary calibrators are expected to be
found nearly anywhere in the sky, so the full-resolution dataset can be shifted
(and averaged) to a number of different regions within the station beam. The
data from the core and remote stations can be used to explore the potential
calibrators/targets in the field at low resolution, and the full-array data
would improve the sensitivity and the resolution of the survey.  Eventually, a
full-resolution image of the whole primary beam can be produced, at the expense
of a very high computational cost.

We found a density of $\sim1$ good calibrator per square degree based on two
fields with Galactic latitudes of $+26.6^{\circ}$ and $+43.4^{\circ} $. However,
we expect less compact sources at lower Galactic latitudes due to interstellar
scattering. The Galactic electron density model NE2001 \citep{cordes02} predicts
an scattering at a galactic latitude of $50^{\circ}$ of almost 100~mas at
150~MHz, which is five times smaller than our resolution. However, the
scattering is about 300~mas, similar to our beamsize, at latitudes of
5--10$^{\circ}$, depending on the longitude. Therefore, observations below a
Galactic latitude of 10$^{\circ}$ are likely to be affected by scattering on the
longest baselines, and the effect should be severe below about 2$^{\circ}$,
especially towards the Galactic Center. Therefore, an accurate analysis of the
area, and a more exhaustive search of calibrators, is required when observing
low Galactic latitudes because the compactness of sources can be significantly
worse than for the cases presented here.

\section{Conclusions}\label{sec:conclusions}

We have observed 630 sources in two fields with the LOFAR international stations
to determine the density of good long-baseline calibrators in the sky.  We have
seen that a number of properties from lower angular resolution data are
correlated with the likelihood of being a suitable calibrator.   High flux
density, a flat low-frequency spectrum, and compactness in the NVSS catalogue
are all useful predictors of calibrator suitability.  The spectral index at
higher frequency, in contrast, is a poor predictor.

The conclusions of this study are:

\begin{enumerate}
\item With a survey speed of $\sim$360 targets per hour in ``snapshot'' survey
mode, identifying the optimal calibrator for an International LOFAR observation
can be cheaply performed before the main observation. 
\item The density of suitable calibrators for International LOFAR observations
in the high band ($\sim$140 MHz) is around 1 per square degree -- high enough
that a suitable calibrator should be found within 1$^{\circ}$ of the target
source virtually anywhere in the sky (excluding regions of high scattering such
as the Galactic plane).
\end{enumerate}

The Multifrequency Snapshot Sky Survey (MSSS) is the first northern-sky LOFAR
imaging survey between 30 and 160~MHz \citep{heald14} with a 90\% completeness
of 100~mJy at 135~MHz. It provides low-resolution images and source catalogues
including detailed spectral information.  At the time of writing this paper the
final catalogue was not available, so it was not included in our analysis. The
MSSS catalogue can be used to improve the selection of potential long baseline
calibrators.

Finally, we anticipate extending this work in the future with observations at
lower frequencies (the LOFAR low band is capable of observing from 15--90 MHz),
although the density of suitable sources is expected to be much lower due to the
lower sensitivity in this frequency range, combined with an even greater impact
of ionospheric conditions.

\appendix
\section{How to plan an International LOFAR observation}\label{apx:plan}

Given the positive results of this project to find and calibrate potential delay
calibrators, we propose the following approach for an International LOFAR
observation:

\begin{enumerate}
\item Identify candidate primary calibrators up to separations of a few degrees 
by using any of the criteria discussed in Sect.~\ref{results};
\item Conduct a short observation in snapshot mode as described in
Sect.~\ref{sec:obs} before the science observation to identify the best primary
calibrator (or calibrators).
\item If required and time permits, follow up with a ``full bandwidth'' snapshot
observation to identify one or more secondary calibrators;
\item Set up the scientific observation to dwell on the field containing the
primary calibrator and the target/secondary calibrator;
\item Include periodic scans (every $\sim$ hour) on a bright Dutch array
calibrator to calibrate the core stations in order to form the tied station.
\item Shift phase centre to primary calibrator, preprocess and obtain delay
solutions as described in this paper, apply them to the unshifted dataset;
\item If a secondary calibrator is to be used and is not yet identified, select
10 minutes of data and perform shift/averaging to candidate secondary calibrator
sources;
\item If secondary calibrator is used: shift and average primary-calibrated
dataset, image and selfcalibrate, apply solutions to the unshifted dataset;
\item Shift and average calibrated dataset, image and (if needed) selfcalibrate
target.
\end{enumerate}

In the near future, the pipeline used for this project will be developed, in
collaboration with the LOFAR operations team, into an expanded form capable of
carrying out the approach described above.  This pipeline will be made available
to all International LOFAR observers, delivering a reduced data volume for
long-baseline observations and enabling calibrated data to be more quickly
produced.

\begin{acknowledgements}

LOFAR, the Low Frequency Array designed and constructed by ASTRON, has
facilities in several countries, that are owned by various parties (each with
their own funding sources), and that are collectively operated by the
International LOFAR Telescope (ILT) foundation under a joint scientific policy.
ATD is supported by a Veni Fellowship from NWO.
ADK acknowledges support from the Australian Research Council Centre of Excellence for All-sky Astrophysics (CAASTRO), through project number CE110001020.
LKM acknowledges financial support from NWO Top LOFAR project, project n. 614.001.006.
CF acknowledges financial support by the {\it ``Agence Nationale de
la Recherche''} through grant ANR-09-JCJC-0001-01.
\end{acknowledgements}


\bibliographystyle{aa} 
\bibliography{biblio} 

\end{document}